\documentclass{emulateapj}

\usepackage{graphicx}
\usepackage{amssymb}
\usepackage[abs]{overpic}

\usepackage{hyperref}
\hypersetup{colorlinks=true,
            citecolor=blue,
            linkcolor=blue}

\newcommand{\tpr}[1]{\textcolor{red}{#1}}    

\def \llabel#1{\label{#1}}

\def\aj{Astron.\ J. }
 
\def\apjl{Astrophys.\ J. }

\def\aap{Astron.\ Astrophys. }

\def\mnras{Mon.\ Not.\ R.\ Astron.\ Soc. }

\def\nat{Nature }

\shorttitle{Spin-orbit coupling for quasi-circular coorbital bodies}
\shortauthors{A.C.M. Correia \& P. Robutel}

\begin{document}


\title{Spin-orbit coupling and chaotic rotation for coorbital \\ bodies in quasi-circular orbits} 


\author{Alexandre C.M. Correia}
\affil{Departamento de F\'isica, I3N, Universidade de Aveiro, Campus de
Santiago, 3810-193 Aveiro, Portugal;}
\affil{Astronomie et Syst\`emes Dynamiques, IMCCE-CNRS UMR8028, 
77 Av. Denfert-Rochereau, 75014 Paris, France}

\author{Philippe Robutel}
\affil{Astronomie et Syst\`emes Dynamiques, IMCCE-CNRS UMR8028, 
77 Av. Denfert-Rochereau, 75014 Paris, France}


\date{2013 November 22}

\begin{abstract}
Coorbital bodies are observed around the Sun sharing their orbits with the planets, but also in some pairs of satellites around Saturn. 
The existence of coorbital planets around other stars has also been proposed.
For close-in planets and satellites, the rotation slowly evolves due to dissipative tidal effects until some kind of equilibrium is reached.
When the orbits are nearly circular, the rotation period is believed to always end synchronous with the orbital period.
Here we demonstrate that for coorbital bodies in quasi-circular orbits, stable non-synchronous rotation is possible for a wide range of mass ratios and body shapes.
We show the existence of an entirely new family of spin-orbit resonances at the frequencies $n\pm k\nu/2$, where $n$ is the orbital mean motion, $\nu$ the orbital libration frequency, and $k$ an integer.
In addition, when the natural rotational libration frequency due to the axial asymmetry, 
$\sigma$, has the same magnitude as $\nu$, 
the rotation becomes chaotic.
Saturn coorbital satellites are synchronous since $\nu\ll\sigma$, but coorbital exoplanets may present non-synchronous or chaotic rotation. 
Our results prove that the spin dynamics of a body cannot be dissociated from its orbital environment.
We further anticipate that a similar mechanism may affect the rotation of bodies in any mean-motion resonance.
\end{abstract}





\section{Introduction}

Coorbital bodies have fascinated astronomers and mathematicians
since \citet{Lagrange_1772} found an equilibrium configuration where three bodies are located at the vertices of an equilateral triangle.
\citet{Gascheau_1843} proved that for a circular motion of the three bodies, the Lagrange equilibrium points were stable under specific conditions fulfilled by the three masses.
In 1906, the first object of this kind was observed \citep{Wolf_1906}, the asteroid Achilles, that shares its orbit with Jupiter around the Sun, leading on average by $60^\circ$. 
At present, more than 4000 coorbital bodies are known in the Solar System\footnote{http://www.minorplanetcenter.net/}, sharing their orbits with the planets. 
More interestingly, pairs of tidally evolved coorbital satellites  were also observed around Saturn in a wide variety of orbital configurations \citep[e.g.][]{Robutel_etal_2012}.  
These objects present very low eccentricities (less than 0.01) and their rotations appear to be synchronous, although there is no confirmation yet \citep{Tiscareno_etal_2009}. 

Tidal dissipation slowly modifies the rotation rate 
of close-in planets and satellites 
\citep[e.g][]{MacDonald_1964,Correia_2009}.
For rigid bodies, when the rotation rate and the mean motion have the same magnitude, 
the dissipative tidal torque may be counterbalanced by the conservative torque due to the axial asymmetry of the inertia ellipsoid.
For eccentric orbits, this conservative torque allows for capture of the spin rate in a half-integer commensurability with the mean motion, usually called spin-orbit resonance \citep{Colombo_1965, Goldreich_Peale_1966, Correia_Laskar_2009}. 
In addition, for very eccentric orbits or large axial asymmetries, 
the rotational libration width of the individual resonances may overlap, and 
the rotation becomes chaotic \citep{Wisdom_etal_1984,Wisdom_1987}.
However, for nearly circular orbits, 
the only possibility for the spin is the synchronous 
resonance \citep[e.g.][]{Goldreich_Peale_1966, Correia_Laskar_2009}. 
Since tidal dissipation simultaneously damps the eccentricity to zero \citep[e.g.][]{Hut_1980, Correia_2009}, all the main satellites in the Solar System are observed in quasi-circular orbits and synchronous rotation. 

Contrarily to the classical two-body problem, 
where circular orbits are unperturbed, in the case of coorbital bodies, the orbits often present long-term librations around the Lagrange equilibrium points.
As a consequence, there is a permanent misalignment of the rotating body long inertia axis from the radius vector to the central body (Fig.~\ref{ref:angles}).
The resulting torque on the rotating body's figure induces some rotational libration \citep{Tiscareno_etal_2009, Robutel_etal_2011, Robutel_etal_2012}. 
The combination of both libration motions (orbital and rotational) may give rise to some unexpected behaviors for the rotation rate. 
In this, paper we investigate all the possibilities for the final rotation of coorbital bodies in quasi-circular orbits.

\section{Model}

\llabel{TheModel}

\begin{figure}
\begin{center}
 \includegraphics[width=0.8\columnwidth]{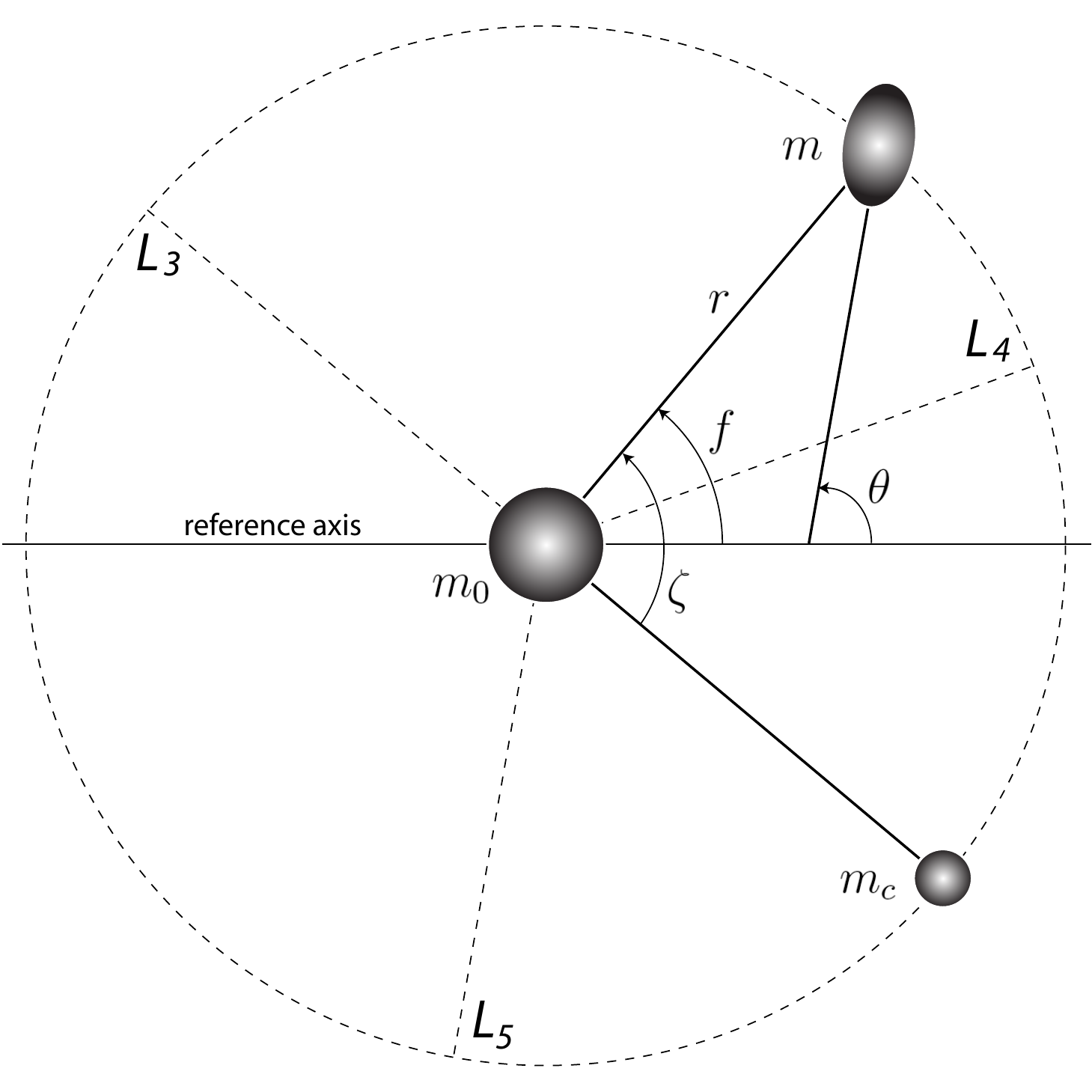}
 \caption{Reference angles for the coorbital system. $m_0$ is the mass of the central body, $m$ the mass of the rotating body, and $m_c$ the mass of the coorbital companion. $\theta$ is the rotation angle of $m$, $r$ its distance to the central body, and $f$ its true anomaly. $\zeta$ is the angle between the directions of $m_c$ and $m$.
 \llabel{ref:angles} }
\end{center}
\end{figure}

Let us denote $m_0$ the mass of the central body, $m$ the mass of the rotating body, and $m_c$ the mass of the coorbital companion (Fig.~\ref{ref:angles}).
We adopt here the theory developed by \citet{Erdi_1977}  adapted to the planetary problem \citep{Robutel_etal_2012}, and limited to the first order in $\mu=(m+m_c)/(m_0+m+m_c)$.
We additionally assume quasi-circular orbits (negligible eccentricity) with average radius $r_0$  for both coorbital bodies. 
The polar coordinates $(r,f)$ for $m$ centered on $m_0$ are given by \citep{Robutel_etal_2012}:
\begin{equation}
r=r_0\left(1-\frac{2\delta}{3n}\dot\zeta\right)\ ,
\label{eq:sol_orb_r}
\end{equation}
\begin{equation}
f=\delta\zeta+nt+f_{0}\ ,
\end{equation}
\begin{equation}
\ddot\zeta=-3\mu n^2\left[1-(2-2\cos\zeta)^{-3/2}\right]\sin\zeta\ ,
\label{eq:sol_orb_zeta} 
\end{equation}%
where $n$ is the orbital mean motion, $\delta=m_c/(m+m_c)$, 
and $f_{0}$ is a constant.
The system can be stable as long as $\mu<0.03812$ \citep{Gascheau_1843, Siegel_Moser_1971}.

\begin{figure}
\begin{center}
 \includegraphics[width=0.9\columnwidth]{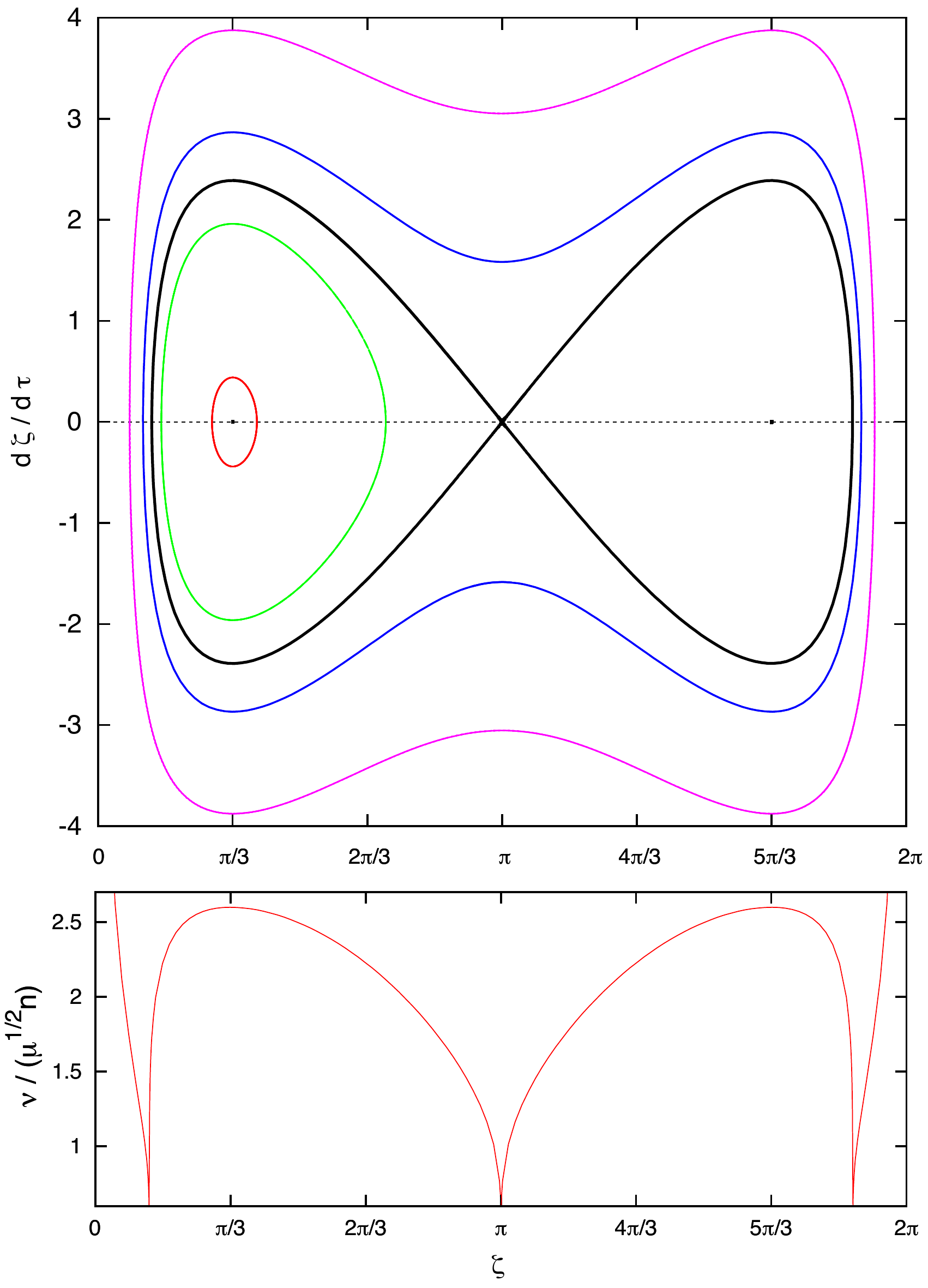} 
 \caption{
Longitudinal variations of coorbital bodies (Eq.~\ref{tau2}) in the plane ($\zeta,d\zeta/d\tau$) (top), and
orbital libration frequency  taken over the dashed line ($\nu$ versus $\zeta$ with $d\zeta/d\tau=0$) (bottom).
The black curve is the separatrix between the tadpole and the horseshoes orbits ($\zeta_\mathrm{min} \approx 24^\circ$). The two tadpole orbits surrounding $L_4$ correspond to $\alpha=10^\circ$ and  $\alpha=50^\circ$, while the two horseshoe orbits are associated to $\alpha=160^\circ$ and $\alpha=166^\circ$.
  \label{fig:orbits}  }
\end{center}
\end{figure}

Equation 
governs the relative angular position of  the coorbitals.
Its solutions, plotted in figure \ref{fig:orbits}, are periodic  
with an associated frequency  $\nu$ of the order of $\sqrt{\mu}n$ \citep{Erdi_1977}.
There are stable equilibria, $L_4$ and $L_5$, for $(\zeta_0,\dot\zeta_0)=(\pm\pi/3,0)$, and unstable equilibrium, $L_3$, for $(\zeta_0,\dot\zeta_0)=(\pi,0)$. 
The trajectories starting with initial conditions $\zeta_0=\pm\pi/3$, ${\vert\dot\zeta}_0\vert<n\sqrt{6\mu}$ describe tadpole orbits around $L_4$ or $L_5$, and those starting from $\zeta_0 = \pm \pi/3$, ${\vert\dot \zeta}_0\vert >n\sqrt{6\mu}$ evolve on horseshoe orbits (Fig.~\ref{fig:orbits}).
The separatrix between these two types of orbits is given by $\zeta_\mathrm{min}\approx24^\circ$ 
 \citep[for more details see][]{Robutel_etal_2012}.

The amplitude of the radial variations is usually very small, 
so that the orbit remains
nearly circular (Eq.~\ref{eq:sol_orb_r}), but the longitudinal half-maximal libration amplitude,
$$
 \alpha\equiv\vert\zeta_\mathrm{max}-\zeta_\mathrm{min}\vert/2
$$
can be very large, depending on the initial conditions (Fig.~\ref{fig:orbits}). 

Let us denote $ A < B < C $ the moments of inertia of the rotating body.
The equation of motion for the rotation angle, $\theta$, is then given 
by \citep[e.g.][]{Murray_Dermott_1999}:
\begin{equation}
\ddot\gamma=-\frac{\sigma^2}{2}\left(\frac{r_0}{r}\right)^3\sin2(\gamma-\delta(\zeta-\zeta_0))
 \ ,\label{eq:rot_ga}
\end{equation}
where 
\begin{equation}
\gamma=\theta-nt-f_{0}-\delta\zeta_0\ ,
\end{equation}
and
\begin{equation}
\sigma=n\sqrt{3\frac{B-A}{C}}\ ,
\end{equation}
which is approximately the frequency for small-amplitude rotational librations.

For the tidal dissipation we adopt a viscous linear model, whose contribution to the rotation is given by \citep{Mignard_1979,Correia_Laskar_2004}:
\begin{equation}
\ddot\gamma=-K\left(\frac{r_0}{r}\right)^6(\dot\gamma-\delta\dot\zeta)
\ ,\label{eq:rot_diss}
\end{equation}
where 
\begin{equation}
K=3n\frac{k_2}{\xi Q}\left(\frac{R}{r_0}\right)^3\left(\frac{m_0}{m}\right) 
\end{equation}
is a dissipation constant, 
with $k_2$ the second Love number, $Q^{-1}\equiv n\Delta t$ the dissipation factor, $\Delta t$ the dissipation time lag, $R$ the radius of the rotating body, and $\xi$ its normalized moment of inertia.

\section{Dynamical Analysis}

For small-amplitude orbital librations ($ \alpha\ll 1$), a simple linear theory can be used to understand the diversity of rotational behaviors.
In this case,
the solution of equation (\ref{eq:sol_orb_zeta}) is given by \citep{Erdi_1977}  
\begin{equation}
\zeta=\zeta_0+ \alpha\sin(\nu t)\ ,
\llabel{linearzeta}
\end{equation}
where $\nu\approx n\sqrt{27\mu}/2$. 
At first order in $\alpha$, 
expression (\ref{eq:rot_ga}) simplifies as:
\begin{equation}
\ddot\gamma=-\frac{\sigma^2}{2}\left[\sin2\gamma+ \alpha^+\sin2(\gamma-\frac{\nu}{2}t)- \alpha^-\sin2(\gamma+\frac{\nu}{2}t)\right] 
\ ,\label{eq:rot_av}
\end{equation}
where $ \alpha^\pm= \alpha(1\pm\nu/n)\delta$.
We then have three main islands of rotational libration,  $\dot\gamma=\dot\theta-n=0,\pm\nu/2$, with half-widths $\sigma$ and $\sigma\sqrt{ \alpha^\pm}$, respectively.
For $\mu\ll 1$ and $m\ll m_c$, we get $ \alpha^\pm= \alpha$.
Therefore, together with the classical synchronous equilibrium at $\dot\theta=n$, there exists two additional possibilities for the spin at the super- and sub-synchronous resonances $\dot\theta=n\pm\nu/2$.

More generally, since $\zeta$ is a periodic function with frequency $\nu$, we can write \citep{Robutel_etal_2011}:
\begin{equation}
\left(\frac{r_0}{r}\right)^3\mathrm{e}^{-\mathrm{i}2\delta\zeta}=\sum_{k\in\mathbb Z}\rho_k\mathrm{e}^{\mathrm{i}(k\nu t+\phi_k)} \ ,
\end{equation}
where $\rho_k$ and $\phi_k$ are the amplitude and the phase shift of each harmonic, that depend on $\alpha$.
Thus, expression (\ref{eq:rot_ga}) becomes:
\begin{equation}
\ddot\gamma=-\frac{\sigma^2}{2}\sum_{k\in\mathbb Z}\rho_k\sin(2\gamma+k\nu t+\phi_k)\ ,\label{eq:rot_gen}
\end{equation}
where the main resonances can be found for $\dot\theta=n\pm k\nu/2$, $k$ being an integer.

\begin{figure}
\begin{center}
\includegraphics[width=0.9\columnwidth]{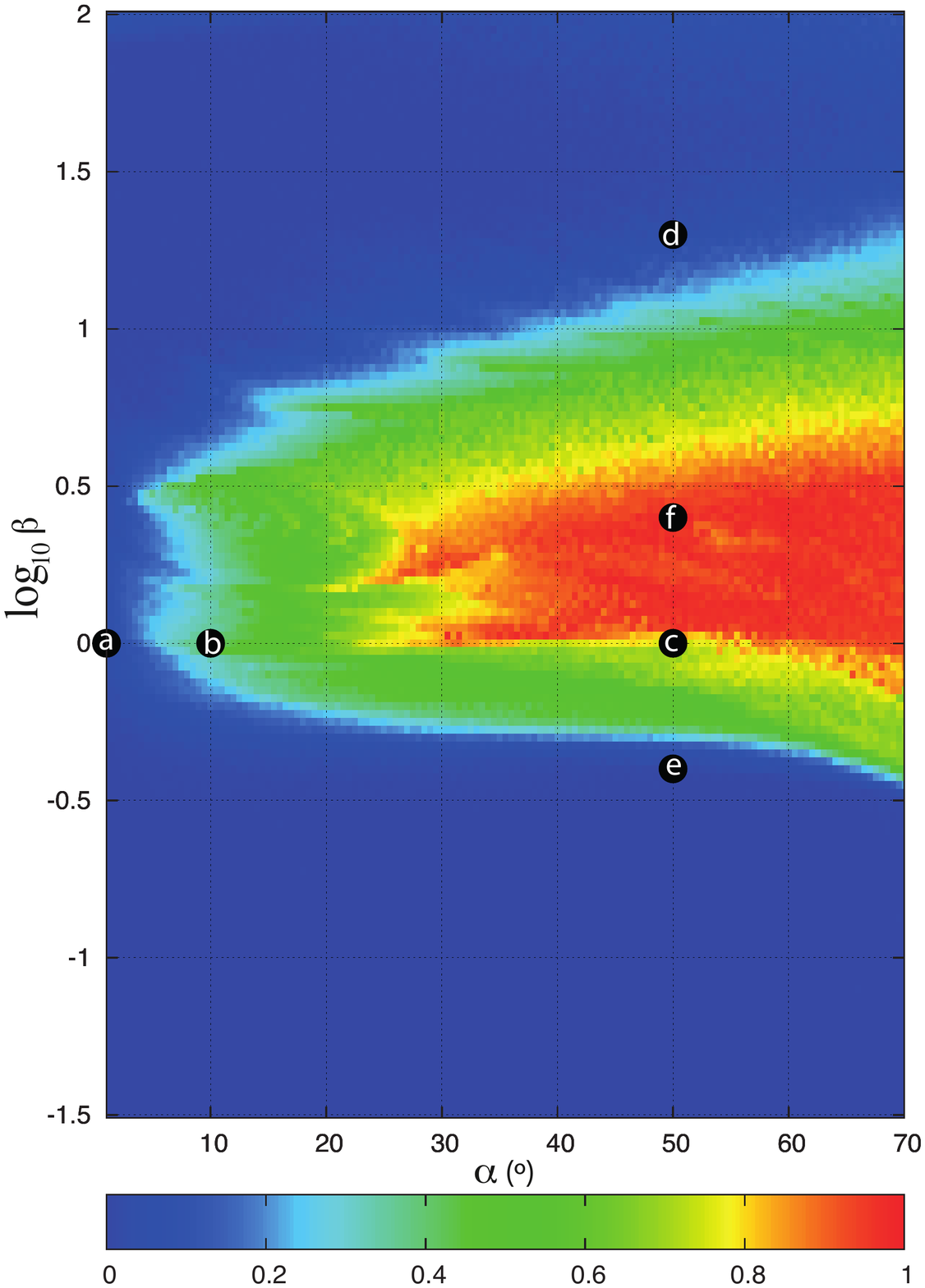} 
 \caption{Stability analysis of the rotation rate
close to the synchronization for  $( \alpha,\log_{10}\beta) \in [0^\circ:70^\circ]\times[-1.5:2]$ (Tadpole orbits). The color index indicates the proportion of chaotic orbits inside the studied domain: from dark blue for fully regular to red for entirely chaotic. 
Above the chaotic region ($\log_{10}\beta>0$) only synchronous rotation is possible, while below it ($\log_{10}\beta<0$) several spin-orbit resonances are possible.  
In this plot we fixed $\mu\simeq m_c/m_0=10^{-3}$ and $m/m_c=10^{-3}$, 
but these values do not significantly affect the results as long as $m/m_c<0.1$ (Fig.~\ref{fig:gloal_T}). 
The black circles correspond to the locations of the six examples presented in Figures~\ref{fig:section} and~\ref{fig:tides}.
    \label{fig:global}  }
\end{center}
\end{figure}

The general problem for the spin-orbit evolution of quasi-circular coorbital bodies is reduced to the analysis of the two frequencies, $\nu$ and $\sigma$, and the amplitude $ \alpha$. 
However, by rescaling the time using $\tau=\sqrt{\mu}nt$, 
we can rewrite  equations (\ref{eq:sol_orb_r}), (\ref{eq:sol_orb_zeta}) and (\ref{eq:rot_ga}) as
\begin{equation}
r=r_0\left(1-\sqrt\mu\frac{2\delta}{3}\frac{d\zeta}{d\tau}\right)\simeq r_0\ ,
\llabel{tau1}
\end{equation}
\begin{equation}
\frac{d^2\zeta}{d\tau^2}=-3\left[1-(2-2\cos\zeta)^{-3/2}\right]\sin\zeta\ ,
\llabel{tau2}
\end{equation}%
\begin{equation}
\frac{d^2\gamma}{d\tau^2}=-\frac{\beta^2}{2}\left(\frac{r_0}{r}\right)^3\sin2(\gamma-\delta(\zeta-\zeta_0))\ ,
\llabel{tau3}
\end{equation}
with
$$
\beta\equiv\sigma/(\sqrt{\mu}n)\sim\sigma/\nu\ .
$$
We see that
the orbital motion is almost independent of $\mu$, since $\mu<0.038$, and that
the rotational motion only depends on $\beta$.
The global dynamics of the spin is then approximately controlled by only two parameters: $ \alpha$ and $\beta$. 

We can perform a stability analysis of $\dot\gamma$ in the plane ($ \alpha,\beta$) to quickly identify the rotational regime for any system of coorbital bodies that is near the synchronous equilibrium.
If isolated, the half-width of the synchronous resonant island in the direction of $d\gamma/d\tau$ is equal to $\beta$.
Thus, for a given $(\alpha,\beta)$
we select $400$ equi-spaced values of $d\gamma/d\tau$ in the interval $[-2\beta:2\beta]$ and fix the initial value of $\gamma$ at the synchronization libration center. 
The corresponding solutions are integrated using the equations (\ref{tau1})-(\ref{tau3}) and their dynamical nature (stable/unstable) is deduced from frequency analysis \citep{Laskar_1990,Laskar_1993PD}, which gives the fraction of chaotic trajectories. 

\begin{figure*}
\begin{center}
\includegraphics[width=0.85\textwidth]{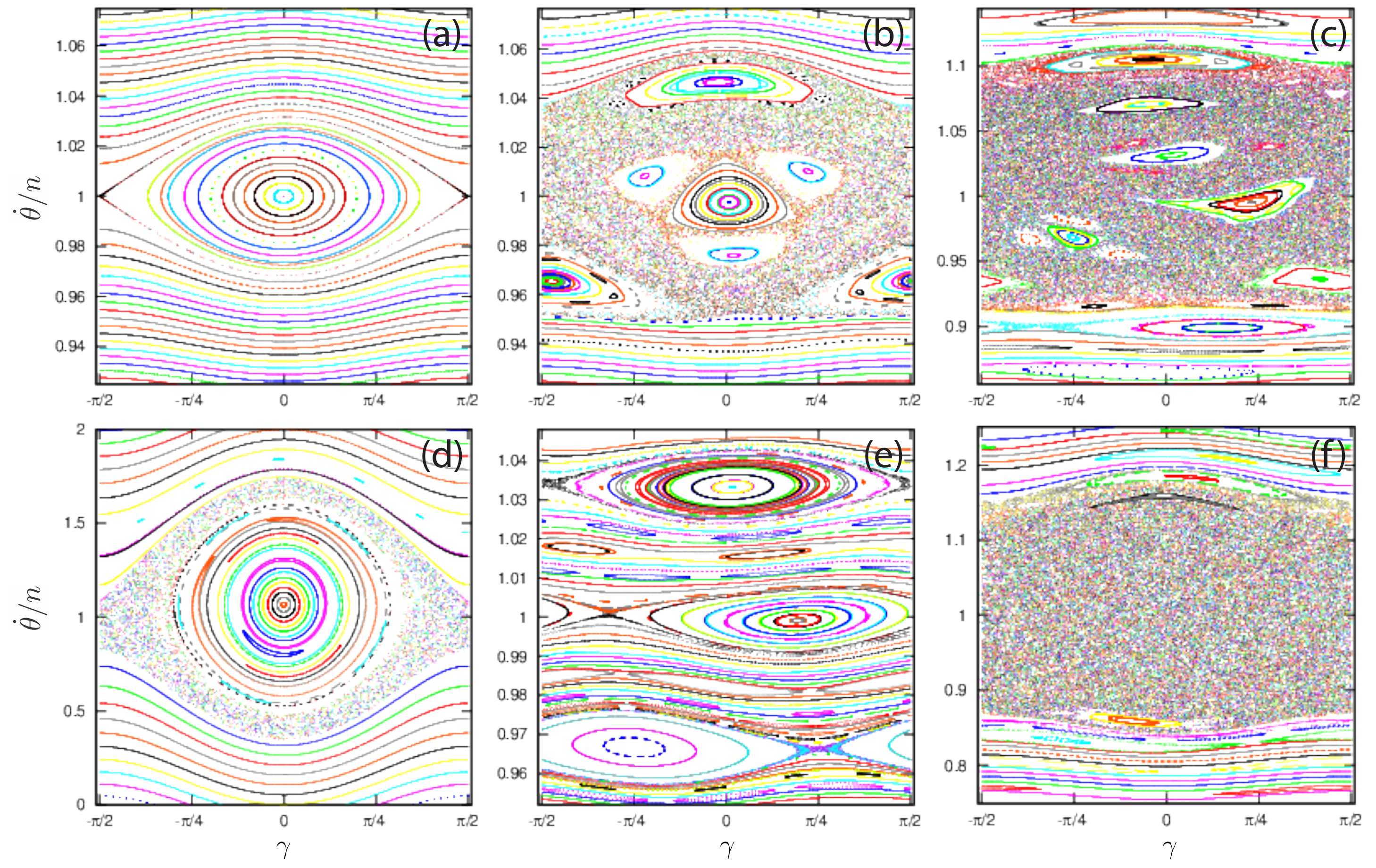} 
  \caption{Poincar\'e surface sections in the plane 
  $(\gamma, \dot\theta/n)$ deduced from the time-$2\pi/\nu$ map of the flow (Eq.~\ref{eq:rot_ga}). The angle $\gamma$ is reduced modulo $\pi$.
  In the top graphs we fix $\beta=1$ and increase the amplitude $ \alpha$: (a) $\alpha = 0^{\circ}$. In this case, the dynamics is the same as a simple pendulum for all $\beta$; 
  (b) $\alpha = 10^{\circ}$. Although the super and sub-synchronous islands are small (Eq.~\ref{eq:rot_av}), the partial overlap of the resonances already generate a significant chaotic region;
  (c) $\alpha = 50^{\circ}$. For large amplitudes, the islands' overlap give rise to a huge unstable region surrounding the synchronization. 
  In the bottom graphs we fix $ \alpha=50^\circ$ and vary $\beta$:
  (d) $\beta=10^{1.3}$. The three islands merged generating a chaotic layer in the neighborhood of the synchronization separatrix, as for the modulated pendulum;
 (e) $\beta=10^{-0.4}$.  The three main resonant islands are isolated, narrow chaotic regions are present along the associated separatrices; 
 (f)  $\beta=10^{0.4}$. Deep inside the chaotic zone, in this case no stable motion is possible.
 \label{fig:section} }
\end{center}
\end{figure*}

In Figure~\ref{fig:global} we show the results for tadpole orbits 
with $\mu\simeq m_c/m_0=10^{-3}$ and $m/m_c=10^{-3}$
(for instance, an Earth-like planet around a Sun-like star with a Jupiter-like coorbital).
The color index indicates the proportion of chaotic orbits inside the studied domain: from dark blue for fully regular to red for entirely chaotic. 
Depending on the $ \alpha$ and  $\beta$ values, the rotation can present a wide variety of behaviors, ranging from non-synchronous equilibria to chaotic motion.
A common way of visualizing and understanding the different regimes is to use Poincar\'e surface sections \citep[e.g.][]{Wisdom_1987,Morbidelli_2002}.
Therefore, we selected a few representative pairs $(\alpha,\beta)$ and plotted the corresponding diagrams in Figure~\ref{fig:section}.

\begin{figure*}
\begin{center}
\includegraphics[width=0.85\textwidth]{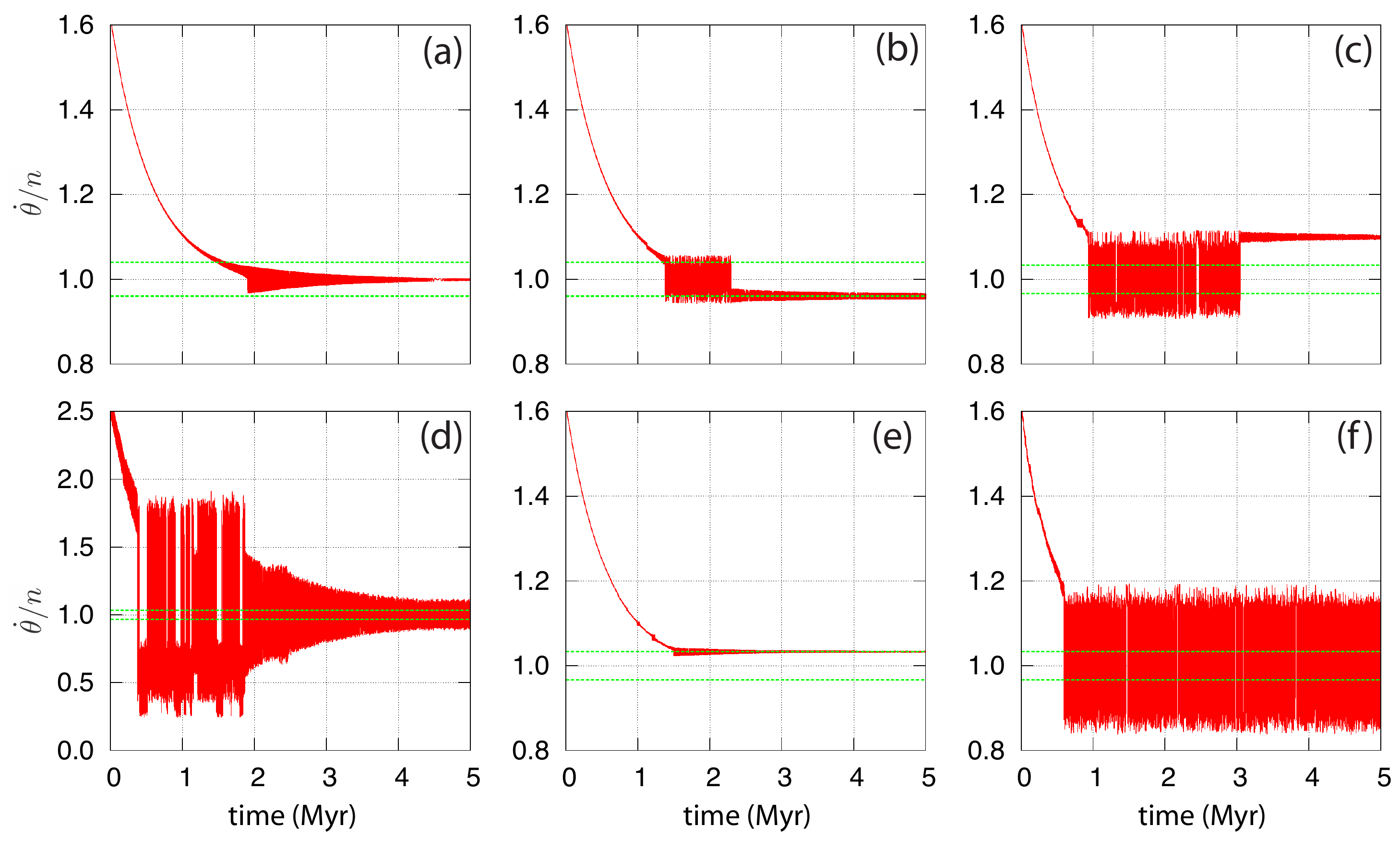} 
  \caption{Examples of the final evolution of the rotation of a coorbital body. We numerical integrate the equations (\ref{eq:sol_orb_r}$-$\ref{eq:rot_ga}) with
$n=17.78$~yr$^{-1}$ (corresponding to $m_0=1~M_\odot$ and $r_0=0.5$~AU),  
  together with tidal dissipation using $K=250$~yr$^{-1}$ (Eq.~\ref{eq:rot_diss}). 
We show an example for each pair ($ \alpha, \beta$) taken from Fig.~\ref{fig:section}.
The initial rotation rate is $\dot\theta/n=2.5$ in (d) and $\dot\theta/n=1.6$ in all the other plots.
The green dotted lines give the position of the super- and sub-synchronous resonances $n\pm\nu/2$ (Eq.~\ref{eq:rot_av}).
   \label{fig:tides}   }
\end{center}
\end{figure*}

\begin{figure*}
\begin{center}
 \includegraphics[width=0.83\textwidth]{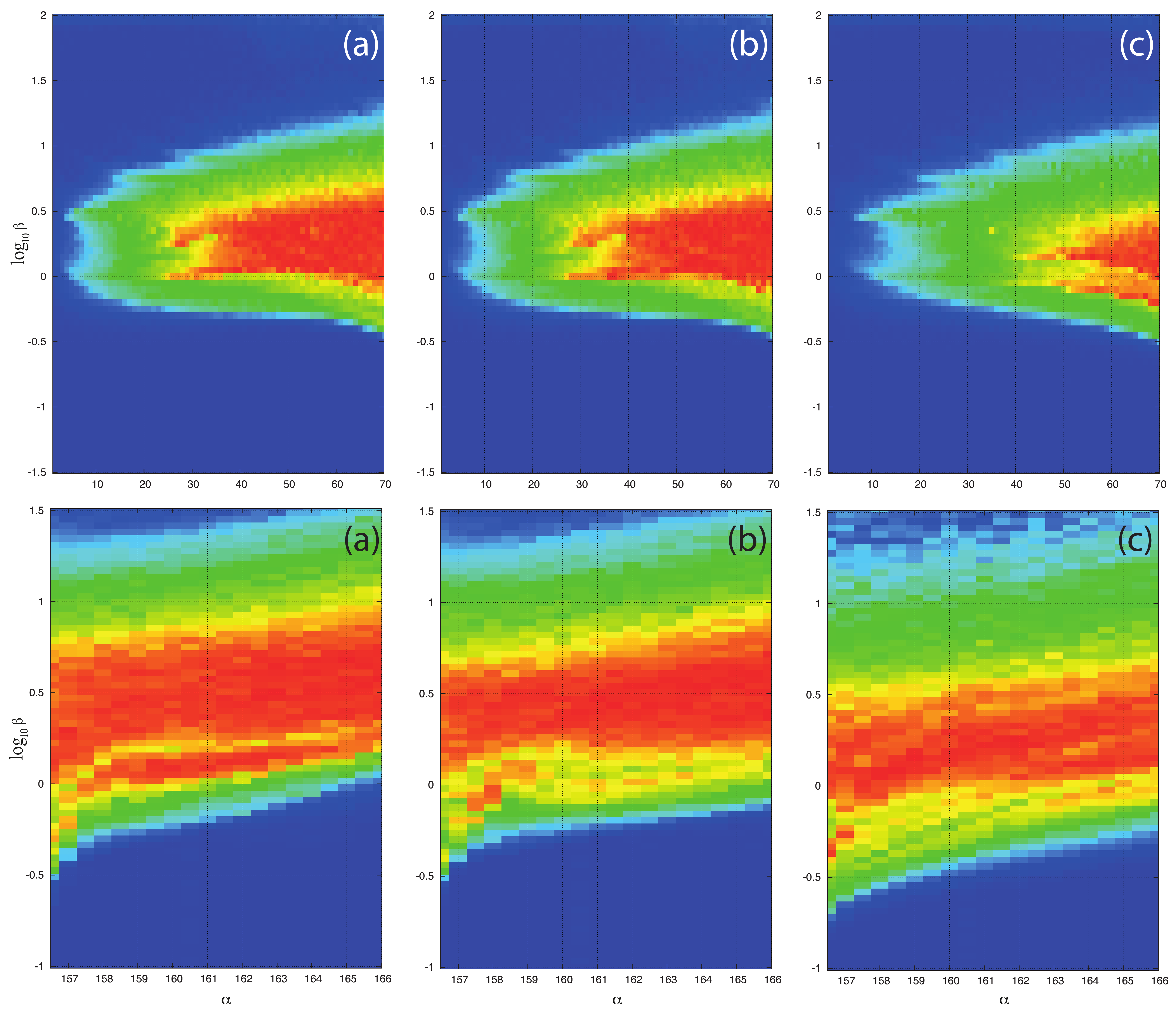} 
 \caption{Same plots as in Figure~1 ($m/m_c=m_c/m_0=10^{-3}$), but for different mass ratios. Top pictures correspond to tadpole-type orbits $( \alpha,\log_{10}\beta) \in [0^\circ:70^\circ]\times[-1.5:2]$, while bottom pictures correspond to horseshoe type orbits $( \alpha,\log_{10}\beta) \in [156^\circ:166^\circ]\times[-1:1.5]$.
We fix $m_c/m_0 = 10^{-6}$ and vary $m/m_c$: (a) $m/m_c=10^{-3}$ ($\delta = 0.999$);  (b) $m/m_c=0.1$ ($\delta = 10/11$); (c) $m/m_c=1$ ($\delta = 1/2$).
  \label{fig:gloal_T}  }
\end{center}
\end{figure*}

For $ \alpha=0$, the coorbital is at equilibrium at a Lagrangian point.
In this particular case, the orbital motion is the same as on an unperturbed circular orbit, so the only possibility for the spin is the synchronous rotation (Fig.~\ref{fig:section}a).
For non-zero amplitude orbital librations $\alpha$,  the equation for the rotation of the body (Eq.~\ref{eq:rot_ga}) can be decomposed in a series of individual resonant terms with frequencies  $n\pm k\nu/2$ (at first order), where $k$ is an integer (Eq.~\ref{eq:rot_gen}).
Each term individually behaves like a pendulum, where the rotation can be trapped.
The amplitude of each term increases with $\alpha$.
For small-amplitude orbital librations ($ \alpha\ll 1$) only the first three terms $k=0,\pm1$ are important (Eq.~\ref{eq:rot_av}).
As $\alpha$ increases, additional spin-orbit equilibria appear and extended chaotic regions 
are possible for the spin.

For $\beta\ll1$, the resonant islands are well separated apart (Fig.~\ref{fig:section}e).
Thus, the rotation can be captured in individual spin-orbit resonances and stay there. 
In the linear approximation (Eq.~\ref{eq:rot_av}), for rotation rates decreasing from higher values, the super-synchronous resonance $\dot\theta=n+\nu/2$ is the most likely possibility (Fig.~\ref{fig:tides}e).
Synchronous rotation is also possible, if the rotation escapes capture in the previous resonance.
For rotation rates increasing from lower values, the sub-synchronous resonance $\dot\theta=n-\nu/2$ is encountered first.
For large-amplitude orbital librations, 
capture in stable higher order spin-orbit resonances is also possible. 

When $\beta\sim1$, 
some individual resonant islands overlap.
Analyzed separately, rotational libration could be expected in neighboring resonances, but
such behavior is not possible and the result is chaotic rotation \citep{Chirikov_1979,Morbidelli_2002} (Fig.~\ref{fig:section}f).
This means that the rotation exhibits random variations in short periods of time.
For moderate chaos, the chaotic region may provide a path into the sub-synchronous resonance (Fig.~\ref{fig:tides}b), or to initially escaped higher-order resonances (Fig.~\ref{fig:tides}c).
In the linear approximation (Eq.~\ref{eq:rot_av}), the chaos is confined within the super- and sub-synchronous resonances ($|\dot\theta-n|\lesssim\sigma(1+2\sqrt \alpha)$) (Fig.~\ref{fig:section}b), but for larger orbital libration amplitudes, the chaotic regions can be extended (Fig.~\ref{fig:section}c and Fig.~\ref{fig:gloal_T}).

For $\beta\gg1$, 
the resonances come closer, and all individual rotational libration widths embrace the synchronous equilibrium.
At this stage, chaotic rotation can still be observed around the separatrix of the synchronous resonance, but the motion is regular nearer the center (Fig.~\ref{fig:section}d).
Therefore, when the rotation of a body is decreasing 
by tidal effect, the rotation will be shortly chaotic during the separatrix transition, but it is ultimately stabilized in the synchronous resonance (Fig.~\ref{fig:tides}d), as if the body evolved in a unperturbed circular orbit (Fig.~\ref{fig:tides}a).

\section{Discussion}

In order to test the reliability of the dynamical picture described in the previous section, 
we have numerical integrated the equations (\ref{eq:sol_orb_r})$-$(\ref{eq:rot_ga}) together with tidal dissipation (Eq.~\ref{eq:rot_diss}) using $K=250$~yr$^{-1}$. 
Dissipation is set to a high value so that we can speed-up the simulations. 
However, lower $K$ values have no impact in the capture scenario \citep[e.g.][]{Henrard_1982}, and the evolution time-scale is roughly proportional to $K^{-1}$.
In Figure~\ref{fig:tides} we show an example for each pair ($ \alpha, \beta$) taken from Figure~\ref{fig:section}.
The initial rotation rate is $\dot\theta/n=2.5$ in Fig.~\ref{fig:tides}d and $\dot\theta/n=1.6$ in all the other plots.
The green dotted lines give the position of the super- and sub-synchronous resonances $n\pm\nu/2$ (Eq.~\ref{eq:rot_av}).
It is interesting to observe that the sub-synchronous resonance can be reached after some wandering in the chaotic zone (Fig.~\ref{fig:tides}b).
Capture in higher order spin-orbit resonances can also occur (Fig.~\ref{fig:tides}c). 

Slightly different initial conditions may lead to totally different final equilibrium configurations.
Different tidal models can also change the individual capture probabilities in resonance \citep{Goldreich_Peale_1966}, but not the global picture described in this paper. 
Instead of using the simplified equations  (\ref{eq:sol_orb_r})$-$(\ref{eq:sol_orb_zeta}),
we also run some simulations integrating direct n-body equations for the orbital motion, but no differences were observed.

We can also test the robustness of the stability analysis diagram shown in Figure~\ref{fig:global} for different mass ratios.
In Figure~\ref{fig:gloal_T} we then fix $m_c/m_0 = 10^{-6}$ and vary $m/m_c$, instead
 of using $m/m_c=m_c/m_0=10^{-3}$ (Fig.~\ref{fig:global}).
We confirm that the impact of $\mu\simeq m_c/m_0$ in the global analysis of the spin is imperceptible (Fig.~\ref{fig:gloal_T}a,b). 
However, for comparable coorbital masses ($m/m_c>0.1$) the chaotic region shrinks (Fig.~\ref{fig:gloal_T}c),
because the resonance width is proportional to $\delta=m_c/(m+m_c)$ (Eq.~\ref{eq:rot_av}). 
For identical coorbital masses ($m=m_c$) the orbital libration amplitude is reduced by one-half, so there is less overlap between neighbor spin-orbit resonances.

In Figure~\ref{fig:gloal_T} we simultaneous plot the stability analysis for tapole-type orbits (top) and horseshoe-type orbits (bottom).
We observe that for horseshoe orbits the same rotational regimes as for tadpole orbits are still present.
However, the chaotic regions are more extended, since the orbital libration amplitude is larger $\alpha\in[156^\circ:166^\circ]$.
Therefore, more harmonics to the expansion of the rotational torque must be taken into account (Eq.~\ref{eq:rot_gen}), increasing the number of relevant spin-orbit resonances and the superposition between them.

In the stability analysis diagrams, $\beta\gg1$ corresponds to large rotational libration widths and/or small mass-ratios (Eq.~\ref{tau3}).
This is exactly the situation for all Saturn's coorbital satellites, due to their prominent ellipsoidal figures ($\sigma/n\sim1$) and tiny mass-ratios ($\mu<10^{-6}$; \citeauthor{Robutel_etal_2012} \citeyear{Robutel_etal_2012}).
As a consequence, the only possibility for these satellites is the synchronous resonance.
However, for close-in exoplanets one can expect smaller axial asymmetries and larger mass-ratios \citep{Laughlin_Chambers_2002, Beauge_etal_2007, Cresswell_Nelson_2009, Giuppone_etal_2010}, that is, smaller $\beta$ values.
For Earth-like planets ($m/m_0\sim10^{-6}$), we have $\sigma/n\sim10^{-3}$  \citep{Yoder_1995cnt}  which gives $\beta\sim10^{-1}$ for a Jupiter-like coorbital ($\mu\sim10^{-3}$), and $\beta\sim1$ for another Earth-like coorbital ($\mu\sim10^{-6}$), respectively.
Thus, Earth-like planets may present non-synchronous or chaotic rotation (Fig.~\ref{fig:tides}), an important point to take into account in future orbital evolution studies \citep[e.g.][]{Rodriguez_etal_2013}, and habitability studies \citep[e.g.][]{Selsis_etal_2007}.

The present work should apply more generally to coorbital bodies in eccentric orbits, for which the classic spin-orbit resonances \citep{Colombo_1965, Goldreich_Peale_1966, Correia_Laskar_2009} will split into several components $(k_1 n\pm k_2\nu)/2$, where $k_1$ and $k_2$ are integers.
It should also apply to other orbital resonant configurations, in particular for low order mean motion resonances.

\acknowledgments

We acknowledge support by PICS05998 France-Portugal program, and Funda\c{c}\~ao para a Ci\^encia e a Tecnologia, Portugal (PEst-C/CTM/LA0025/2011).



\begin{thebibliography}{0}%
\makeatletter
\providecommand \@ifxundefined [1]{%
 \@ifx{#1\undefined}
}%
\providecommand \@ifnum [1]{%
 \ifnum #1\expandafter \@firstoftwo
 \else \expandafter \@secondoftwo
 \fi
}%
\providecommand \@ifx [1]{%
 \ifx #1\expandafter \@firstoftwo
 \else \expandafter \@secondoftwo
 \fi
}%
\providecommand \natexlab [1]{#1}%
\providecommand \enquote  [1]{``#1''}%
\providecommand \bibnamefont  [1]{#1}%
\providecommand \bibfnamefont [1]{#1}%
\providecommand \citenamefont [1]{#1}%
\providecommand \href@noop [0]{\@secondoftwo}%
\providecommand \href [0]{\begingroup \@sanitize@url \@href}%
\providecommand \@href[1]{\@@startlink{#1}\@@href}%
\providecommand \@@href[1]{\endgroup#1\@@endlink}%
\providecommand \@sanitize@url [0]{\catcode `\\12\catcode `\$12\catcode
  `\&12\catcode `\#12\catcode `\^12\catcode `\_12\catcode `\%12\relax}%
\providecommand \@@startlink[1]{}%
\providecommand \@@endlink[0]{}%
\providecommand \url  [0]{\begingroup\@sanitize@url \@url }%
\providecommand \@url [1]{\endgroup\@href {#1}{\urlprefix }}%
\providecommand \urlprefix  [0]{URL }%
\providecommand \Eprint [0]{\href }%
\providecommand \doibase [0]{http://dx.doi.org/}%
\providecommand \selectlanguage [0]{\@gobble}%
\providecommand \bibinfo  [0]{\@secondoftwo}%
\providecommand \bibfield  [0]{\@secondoftwo}%
\providecommand \translation [1]{[#1]}%
\providecommand \BibitemOpen [0]{}%
\providecommand \bibitemStop [0]{}%
\providecommand \bibitemNoStop [0]{.\EOS\space}%
\providecommand \EOS [0]{\spacefactor3000\relax}%
\providecommand \BibitemShut  [1]{\csname bibitem#1\endcsname}%
\let\auto@bib@innerbib\@empty
\end{thebibliography}%


\begin{thebibliography}{}

\bibitem[\protect\citeauthoryear{{Beaug{\'e}} et~al.}{{Beaug{\'e}}
  et~al.}{2007}]{Beauge_etal_2007}
{Beaug{\'e}}, C., {S{\'a}ndor}, Z., {{\'E}rdi}, B.,  \& {S{\"u}li}, {\'A}.
  2007, \aap, 463, 359

\bibitem[\protect\citeauthoryear{{Chirikov}}{{Chirikov}}{1979}]{Chirikov_1979}
{Chirikov}, B.~V. 1979, Physics Reports, 52, 263

\bibitem[\protect\citeauthoryear{{Colombo}}{{Colombo}}{1965}]{Colombo_1965}
{Colombo}, G. 1965, \nat, 208, 575

\bibitem[\protect\citeauthoryear{{Correia}}{{Correia}}{2009}]{Correia_2009}
{Correia}, A.~C.~M. 2009, \apjl, 704, L1

\bibitem[\protect\citeauthoryear{{Correia} \& {Laskar}}{{Correia} \&
  {Laskar}}{2004}]{Correia_Laskar_2004}
{Correia}, A.~C.~M.,  \& {Laskar}, J. 2004, \nat, 429, 848

\bibitem[\protect\citeauthoryear{{Correia} \& {Laskar}}{{Correia} \&
  {Laskar}}{2009}]{Correia_Laskar_2009}
{Correia}, A.~C.~M.,  \& {Laskar}, J. 2009, Icarus, 201, 1

\bibitem[\protect\citeauthoryear{{Cresswell} \& {Nelson}}{{Cresswell} \&
  {Nelson}}{2009}]{Cresswell_Nelson_2009}
{Cresswell}, P.,  \& {Nelson}, R.~P. 2009, \aap, 493, 1141

\bibitem[\protect\citeauthoryear{{{\'E}rdi}}{{{\'E}rdi}}{1977}]{Erdi_1977}
{{\'E}rdi}, B. 1977, Celestial Mechanics, 15, 367

\bibitem[\protect\citeauthoryear{{Gascheau}}{{Gascheau}}{1843}]{Gascheau_1843}
{Gascheau}, G. 1843, C. R. Acad. Sci. Paris, 16, 393

\bibitem[\protect\citeauthoryear{{Giuppone} et~al.}{{Giuppone}
  et~al.}{2010}]{Giuppone_etal_2010}
{Giuppone}, C.~A., {Beaug{\'e}}, C., {Michtchenko}, T.~A.,  \& {Ferraz-Mello},
  S. 2010, \mnras, 407, 390

\bibitem[\protect\citeauthoryear{{Goldreich} \& {Peale}}{{Goldreich} \&
  {Peale}}{1966}]{Goldreich_Peale_1966}
{Goldreich}, P.,  \& {Peale}, S. 1966, \aj, 71, 425

\bibitem[\protect\citeauthoryear{{Henrard}}{{Henrard}}{1982}]{Henrard_1982}
{Henrard}, J. 1982, Celestial Mechanics, 27, 3

\bibitem[\protect\citeauthoryear{{Hut}}{{Hut}}{1980}]{Hut_1980}
{Hut}, P. 1980, \aap, 92, 167

\bibitem[\protect\citeauthoryear{{Lagrange}}{{Lagrange}}{1772}]{Lagrange_1772}
{Lagrange}, J.~J. 1772, \OE uvres Compl\`etes VI, 272 (Paris: Gauthier-Villars),
(1869)

\bibitem[\protect\citeauthoryear{{Laskar}}{{Laskar}}{1990}]{Laskar_1990}
{Laskar}, J. 1990, Icarus, 88, 266

\bibitem[\protect\citeauthoryear{{Laskar}}{{Laskar}}{1993}]{Laskar_1993PD}
{Laskar}, J. 1993, Physica D, 67, 257

\bibitem[\protect\citeauthoryear{{Laughlin} \& {Chambers}}{{Laughlin} \&
  {Chambers}}{2002}]{Laughlin_Chambers_2002}
{Laughlin}, G.,  \& {Chambers}, J.~E. 2002, \aj, 124, 592

\bibitem[\protect\citeauthoryear{{MacDonald}}{{MacDonald}}{1964}]{MacDonald_1964}
{MacDonald}, G.~J.~F. 1964, Revs. Geophys., 2, 467

\bibitem[\protect\citeauthoryear{{Mignard}}{{Mignard}}{1979}]{Mignard_1979}
{Mignard}, F. 1979, Moon and Planets, 20, 301

\bibitem[\protect\citeauthoryear{{Morbidelli}}{{Morbidelli}}{2002}]{Morbidelli_2002}
{Morbidelli}, A. 2002, {Modern celestial mechanics : aspects of solar system
  dynamics} (London: Taylor \& Francis)

\bibitem[\protect\citeauthoryear{{Murray} \& {Dermott}}{{Murray} \&
  {Dermott}}{1999}]{Murray_Dermott_1999}
{Murray}, C.~D.,  \& {Dermott}, S.~F. 1999, {Solar System Dynamics} (Cambridge
  University Press)

\bibitem[\protect\citeauthoryear{{Robutel}, {Rambaux}, \&
  {Castillo-Rogez}}{{Robutel} et~al.}{2011}]{Robutel_etal_2011}
{Robutel}, P., {Rambaux}, N.,  \& {Castillo-Rogez}, J. 2011, Icarus, 211, 758

\bibitem[\protect\citeauthoryear{{Robutel}, {Rambaux}, \& {El
  Moutamid}}{{Robutel} et~al.}{2012}]{Robutel_etal_2012}
{Robutel}, P., {Rambaux}, N.,  \& {El Moutamid}, M. 2012, Celestial Mechanics
  and Dynamical Astronomy, 113, 1

\bibitem[\protect\citeauthoryear{{Rodr{\'{\i}}guez}, {Giuppone}, \&
  {Michtchenko}}{{Rodr{\'{\i}}guez} et~al.}{2013}]{Rodriguez_etal_2013}
{Rodr{\'{\i}}guez}, A., {Giuppone}, C.~A.,  \& {Michtchenko}, T.~A. 2013,
  Celestial Mechanics and Dynamical Astronomy, 117, 59

\bibitem[\protect\citeauthoryear{{Selsis} et~al.}{{Selsis}
  et~al.}{2007}]{Selsis_etal_2007}
{Selsis}, F., {Kasting}, J., {Levrard}, B., {Paillet}, J., {Ribas}, I.,  \&
  {Delfosse}, X. 2007, \aap, 476, 1373

\bibitem[\protect\citeauthoryear{{Siegel} \& {Moser}}{{Siegel} \&
  {Moser}}{1971}]{Siegel_Moser_1971}
{Siegel}, C.~L.,  \& {Moser}, J. 1971, {Lectures on celestial mechanics}
  (Berlin: Springer), 13

\bibitem[\protect\citeauthoryear{{Tiscareno}, {Thomas}, \& {Burns}}{{Tiscareno}
  et~al.}{2009}]{Tiscareno_etal_2009}
{Tiscareno}, M.~S., {Thomas}, P.~C.,  \& {Burns}, J.~A. 2009, Icarus, 204, 254

\bibitem[\protect\citeauthoryear{{Wisdom}}{{Wisdom}}{1987}]{Wisdom_1987}
{Wisdom}, J. 1987, \aj, 94, 1350

\bibitem[\protect\citeauthoryear{{Wisdom}, {Peale}, \& {Mignard}}{{Wisdom}
  et~al.}{1984}]{Wisdom_etal_1984}
{Wisdom}, J., {Peale}, S.~J.,  \& {Mignard}, F. 1984, Icarus, 58, 137

\bibitem[\protect\citeauthoryear{{Wolf}}{{Wolf}}{1906}]{Wolf_1906}
{Wolf}, M. 1906, Astronomische Nachrichten, 170, 353

\bibitem[\protect\citeauthoryear{{Yoder}}{{Yoder}}{1995}]{Yoder_1995cnt}
{Yoder}, C.~F. 1995, in Global Earth Physics: A Handbook of Physical Constants,
ed. T. J. Ahrens (Washington, DC: American Geophysical Union), 1
\end{thebibliography}

\end{document}